\documentclass[5p,times,number,sort&compress]{elsarticle}

\usepackage[utf8]{inputenc}
\usepackage{url}
\usepackage{listings}
\usepackage{subcaption}
\usepackage{bbding}
\usepackage{pifont}
\usepackage{array}
\usepackage{hyperref}
\usepackage{xcolor}
\usepackage{booktabs}
\usepackage{amsmath}
\usepackage[english]{babel}

\usepackage{caption}
\usepackage{float}
\usepackage{newfloat}
\DeclareFloatingEnvironment[
   fileext=lst,
   name=Listing
]{listing}
\floatname{listing}{Listing}

\newcommand\code[1]{{\small{\texttt{#1}}}}
\newcommand\fncode[1]{{\scriptsize{\texttt{#1}}}}

\usepackage{xspace}
\usepackage{etoolbox} 
\newcommand{\CC}{C\nolinebreak\hspace{-.05em}\raisebox{.1ex}{\small +}\nolinebreak\hspace{-.05em}\raisebox{.1ex}{\small +}\xspace}

\robustify{\CC}
\newcommand{\ompi}{Open~MPI\xspace}
\newcommand{\parsec}{PaRSEC}

\graphicspath{{figures/}}

\begin{document}

\begin{frontmatter}

\title{%
Callback-based Completion Notification using MPI Continuations\tnoteref{t1}
}
\tnotetext[t1]{Published article in Parallel Computing: \url{https://doi.org/10.1016/j.parco.2021.102793}}

\author[1,2]{Joseph Schuchart\corref{cor1}}
\ead{schuchart@icl.utk.edu}
\author[3]{Philipp Samfass}
\ead{samfass@in.tum.de}

\author[1]{Christoph Niethammer}
\ead{niethammer@hlrs.de}

\author[1]{Jos\'{e} Gracia}
\ead{gracia@hlrs.de}

\author[2]{George Bosilca}
\ead{bosilca@icl.utk.edu}


\affiliation[1]{
  organization={High-Performance Computing Center Stuttgart (HLRS)},
  addressline={Nobelstraße 19},
  postcode={70597},
  city={Stuttgart},
  country={Germany}}

\affiliation[2]{
  organization={Innovative Computing Laboratory (ICL), University of Tennessee Knoxville (UTK)},
  addressline={1122 Volunteer Blvd},
  city={Knoxville, TN},
  postcode={37996},
  country={U.S.A.}}
  
 \affiliation[3] {
   organization={Department of Informatics, Technical University of Munich, Garching, Germany},
   addressline = {Boltzmannstr. 3},
   postcode = {85748},
   city = {Garching (Munich)},
   country = {Germany}}
  
\cortext[cor1]{Corresponding author}

\begin{abstract}
Asynchronous programming models (APM) are gaining more and more traction, allowing applications to expose the available concurrency to a runtime system tasked with coordinating the execution.
While MPI has long provided support for multi-threaded communication and non-blocking operations, it falls short of adequately supporting APMs as correctly and efficiently handling MPI communication in different models is still a challenge.
We have previously proposed an extension to the MPI standard providing operation completion notifications using callbacks, so-called MPI Continuations.
This interface is flexible enough to accommodate a wide range of different APMs.

In this paper, we present an extension to the previously described interface that allows for finer control of the behavior of the MPI Continuations interface.
We then present some of our first experiences in using the interface in the context of different applications, including the NAS parallel benchmarks, the \parsec{} task-based runtime system, and a load-balancing scheme within an adaptive mesh refinement solver called ExaHyPE.
We show that the interface, implemented inside \ompi, enables low-latency, high-throughput completion notifications that outperform solutions implemented in the application space.

\end{abstract}


\end{frontmatter}


\section{Background and Motivation}

Asynchronous (task-based) programming models are gaining more and more traction, promising to help users better utilize ubiquitous multi- and many-core systems by exposing all available concurrency to a runtime system.
Applications are expressed in terms of work-packages (tasks) with well-defined inputs and outputs, which guide the runtime scheduler in determining the correct execution order of tasks.
A wide range of task-based programming models have been developed, ranging from node-local approaches such as OpenMP~\cite{openmp5.0} and OmpSs~\cite{ompss2011} to distributed systems such as HPX~\cite{hpx:2014}, StarPU~\cite{Augonnet:2011}, DASH~\cite{Schuchart:2019:GTD}, and PaRSEC~\cite{DAGuE:2011}.
All of them have in common that a set of tasks is scheduled for execution onto a set of worker threads based on constraints provided by the user, either in the form of dependencies or dataflow expressions.

At the same time, MPI is still the dominant interface for inter-process communication in parallel applications~\cite{Berholdt:2018:MPIUsage}, providing blocking and non-blocking point-to-point and collective operations as well as I/O capabilities on top of elaborate datatype and process group abstractions~\cite{mpi3.1}.
Recent years have seen significant improvements to the way MPI implementations handle communication by multiple threads of execution in parallel~\cite{Patinyasakdikul:2019:GMT,Hjelm:2018:IMM}.

However, task-based programming models pose new challenges to the way applications interact with MPI.
While MPI provides non-blocking communications that can be used to hide communication latencies, it is left to the application layer to ensure that all necessary communication operations are eventually initiated to avoid deadlocks, that in-flight communication operations are eventually completed, and that the interactions between communicating tasks are correctly handled.
Higher-level distributed tasking approaches such as PaRSEC or HPX commonly track the state of active communication operations and regularly test for completion before acting upon such a change in state.
With node-local programming models such as OpenMP, this tracking of active communication operations is left to the application layer.
Unfortunately, the simple approach of test-yield cycles inside a task does not provide optimal performance due to CPU cycles being wasted on testing and (in the case of OpenMP) may not even be portable~\cite{Schuchart:2018:TIT}.

MPI is thus currently not well equipped to support users and developers of asynchronous programming models, which (among other factors) has prompted some higher level runtimes to move away from MPI towards more low-level, asynchronous APIs~\cite{Daliss:2019:FPD}.

Different approaches have been proposed to mitigate this burden, including Task-Aware MPI (TAMPI)~\cite{Sala:2019:TAMPI} and a tight integration of fiber and tasking libraries with MPI implementations~\cite{Mercier:2009:NMA,Lu:2015:MUO,Sala:2018:IIM}.
In a previous paper, we have argued that such an integration, while seemingly attractive due to its ease-of-use, fosters the development of a non-portable application eco-system~\cite{Schuchart:2020:Fibers}. 
In the same paper, we have proposed an extension to the MPI interface, called \emph{MPI Continuations}, that is designed to loosely couple asynchronous programming models with MPI.
In the current paper, we update and extend the description of this interface (Sections~\ref{sec:continuations} and~\ref{sec:rationale}). 
We compare our approach to the MPI\_T interface, a general-purpose API using callbacks for event notification that will be part of the upcoming MPI 4.0 standard (\autoref{sec:mpi_t}).
We will demonstrate the integration of continuations with the PaRSEC runtime and with a adaptive load balancing scheme based on Intel TBB (\autoref{sec:evaluation}).

\section{MPI Continuations: API Overview}
\label{sec:continuations}

Continuations are a concept for structuring the execution of different parts of an application's code, dating back to research on Algol60~\cite{Friedman:1984:PWC,Reynolds;1993:TDC}.
They have recently been proposed to the \CC standard in the form of \code{std::future::then} to coordinate asynchronous activities~\cite{CPP:2014:stdfuture}.
Continuations consist of a \emph{body} (the code to execute) and a \emph{context} (some state passed to the body) on which to operate.

A similar mechanism for MPI can be devised that allows a callback to be attached to an MPI request, which will be invoked once the operation represented by the request has completed.
This establishes a notification scheme to be used by applications to timely react to the completion of operations, e.g., to wake up a blocked thread or release the dependencies of a detached task, all while relieving applications of managing MPI request objects themselves.

We propose a set of functions to set up continuations for active MPI operations, which may be invoked by application threads during calls to communication-related MPI functions or by an implementation-internal progress thread (if available) once all relevant operations are found to have completed.
The body of a continuation may call any MPI library function and thus start new MPI operations in response to the completion of previous ones, e.g., in order to re-post a completed receive.

The MPI Continuations API consists of two parts.
First, \emph{continuation requests} are used to aggregate and progress continuations.
Second, continuations are attached to active MPI operations and registered with continuation requests for tracking.

\begin{listing}
\begin{lstlisting}
/* The continuation callback function definition */
typedef void (MPIX_Continue_cb_function)(
  MPI_Status *statuses,
  void       *cb_data);

/* Create a continuation request using the
   provided info controls */
int MPIX_Continue_init(
  MPI_Request  *cont_req,
  MPI_Info      info);
  
/* Attach a continuation to an active operation represen-
 * ted by op_request. Upon completion of the operation,
 * the callback cb will be invoked and passed status and
 * cb_data as arguments. The status object will be set
 * before the continuation is invoked. If the operation
 * has completed already the continuation will not 
 * be attached or invoked, flag will be set to 1, and
 * the status will be set before return. */
int MPIX_Continue(
  MPI_Request               *op_request,
  int                       *flag,
  MPIX_Continue_cb_function *cb,
  void                      *cb_data,
  MPI_Status                *status,
  MPI_Request                cont_req);
  
/* Similar to the above except that the continuation
 * is invoked once all op_requests have completed. */
int MPIX_Continueall(
  int                        count,
  MPI_Request                op_requests[],
  int                       *flag,
  MPIX_Continue_cb_function *cb,
  void                      *cb_data,
  MPI_Status                 statuses[],
  MPI_Request                cont_req);
\end{lstlisting}
\caption{MPI Continuation interface.}
\label{lst:mpi_continue_fn}
\end{listing}

\subsection{Continuation Requests}

\emph{Continuation requests} (CR) are a form of persistent requests that are created through a call to \code{MPIX\_Continue\_init} and released eventually using \code{MPI\_Request\_free}.
A CR tracks a set of active continuations that are registered with it.
The set grows upon registration of a new continuation and shrinks once a continuation has been executed.
A call to \code{MPI\_Test} on a CR returns \code{flag\,==\,1} if no active continuations are registered.
Conversely, a call to \code{MPI\_Wait} blocks until all registered continuations have completed.
Further details are discussed in \autoref{sec:rationale}.

Continuation requests serve two main purposes.
First, they aggregate continuations attached to active operations and enable testing and waiting for their completion.
Second, by calling \code{MPI\_Test} on a CR, applications can progress outstanding operations and continuations if no MPI-internal progress  mechanism exists to process them asynchronously.
See \autoref{sec:rationale_progress} for a discussion on issues related to progress.

The \code{info} argument to \code{MPIX\_Continue\_init} may be used to control certain aspects of the continuations.
A list of proposed keys will be discussed in \autoref{sec:continue_info}.

\subsection{Registration of Continuations}
A continuation is attached to one or several active operations and registered with the \emph{continuation request} for tracking.
A call to \code{MPIX\_Continue} attaches a continuation to a single operation request while the use of \code{MPIX\_Continueall} sets up the continuation to be invoked once \emph{all} operations represented by the provided requests have completed.
As shown in \autoref{lst:mpi_continue_fn}, the continuation is represented through a callback function with the signature of \code{MPIX\_Continue\_cb\_function} (provided as function pointer \code{cb}) and a context (\code{cb\_data}).
Together with the pointer value provided for the \code{status} or \code{statuses} argument, the \code{cb\_data} will be passed to \code{cb} upon invocation.

Upon return, all provided non-persistent requests are set to \code{MPI\_REQUEST\_NULL}, effectively releasing them back to MPI.\footnote{This has been a deliberate design decision. Otherwise, the release of a request object inside the continuation or the MPI library would have to be synchronized with operations on it outside of the continuation. We thus avoid this source of errors.}
No copy of the requests should be used to cancel or test/wait for the completion of the associated operations.
Consequently, only one continuation may be attached to an operation request.
In contrast, persistent requests may still be canceled, tested, and waited for.

Similar to \code{MPI\_Test}, \code{flag} shall be set to \code{1} if all operations are already complete.
In that case, the continuation callback shall \emph{not} be invoked by MPI, leaving it to the application to handle immediate completion (see \autoref{sec:restrictions}).
Otherwise, \code{flag} is set to zero.

Unless  \code{MPI\_STATUS\_IGNORE} or \code{MPI\_STATUSES\_IGNORE} is provided, the status object(s) will be set before the continuation is invoked (or before returning from \code{MPIX\_Continue[all]} in case all operations have completed already).
The status objects should be allocated by the application in a memory location that remains valid until the callback has been invoked, e.g., on the stack of a blocked thread or fiber or in memory allocated on the heap that is released inside the continuation callback.

\subsection{Usage in a Simple Example}
\label{sec:simple_example}

\begin{listing}[t]
\begin{lstlisting}[numbers=right, breaklines=true, postbreak=\mbox{\textcolor{red}{$\hookrightarrow$}\space}]
bool poll_mpi(MPI_Request *cont_req) { |\label{line:poll_mpi1}|
  int flag; /* result stored in flag ignored here */
  MPI_Test(&flag, cont_req, MPI_STATUS_IGNORE);
  return false; /* we should be called again */
}    |\label{line:poll_mpi2}|
void continue_cb(MPI_Status *status, void *task) { |\label{line:continue_cb1}|
  /* release the task's dependencies */
  nanos6_decrease_task_event_counter(task, 1);
} |\label{line:continue_cb2}|
void solve(int NT, int num_fields, 
           field_t *fields[num_fields]) {
  /* create continuation request */
  MPI_Request cont_req;
  MPIX_Continue_init(&cont_req, MPI_INFO_NULL);
  /* register polling service */
  nanos6_register_polling_service( |\label{line:poll_mpi3}|
      "MPI", &poll_mpi, &cont_req);

  for (int timestep = 0; timestep < NT; timestep++) {
    for (int i = 0; i < num_fields; ++i) {
      #pragma oss task depend(out: fields[i])   |\label{line:task1}|
      {
        /* unpack recv buffer, compute and pack
         * send buffer */
        integrate_halo(fields[i]);        |\label{line:task3}|
        solve(fields[i]);
        save_boundary(fields[i]);            |\label{line:task4}|
        
        /* start send and recv */
        MPI_Request reqs[2];     |\label{line:continue_reg1}|
        MPI_Isend(fields[i]->sendbuf, ..., &reqs[0]);
        MPI_Irecv(fields[i]->recvbuf, ..., &reqs[1]);
        
        /* detach task if requests are active */
        void* task = nanos6_get_current_event_counter();
        nanos6_increase_current_task_event_counter(
            task, 1);
        /* attach continuation */
        int flag;
        MPIX_Continueall(
            2, reqs, &flag, &continue_cb, task,
            MPI_STATUSES_IGNORE, cont_req);
        if (flag)
          nanos6_decrease_task_event_counter(task, 1);  |\label{line:continue_reg2}\label{line:task2}|
  } } }
  /* wait for all tasks to complete and tear down */
  #pragma oss taskwait
  nanos6_unregister_polling_service(
      "MPI", &poll_mpi, NULL);
  MPI_Request_free(&cont_req);
}
\end{lstlisting}
\caption{Simplified example using MPI Continuations in an iterative solver. Tasks are created for each field per timestep. Communication is initiated and tasks are detached. As soon as the communication completes, the task's dependencies are released and the field's next iteration may be scheduled.}
\label{lst:mpi_continue_ex}
\end{listing}

The example provided in \autoref{lst:mpi_continue_ex} uses MPI Continuations to release detached tasks in an OmpSs-2~\cite{BSC:2019:OmpSs2} application once communication operations have completed.
Looping over all fields of a local domain in each timestep, one task is created per field, which carries an output dependency on its \code{field} object (Line~\ref{line:task1}).
Inside each task, the received boundary data from the previous timestep is incorporated and a solver is applied to the field (potentially with nested tasks) before the local boundary is packed for sending (Lines~\ref{line:task3}--\ref{line:task4}).

Both the send and receive operations are initiated and a continuation is attached using \code{MPIX\_Continueall} (Lines~\ref{line:continue_reg1}--\ref{line:continue_reg2}).
OmpSs-2 offers direct access to API functions of the underlying \code{nanos6} runtime.
In this case, the event counter for the current task is retrieved, increased, and passed as the context of the continuation.
If all operation completed immediately, the event counter is decremented again.\footnote{The event counter has to be incremented first as the OmpSs-2 specification mandates that ``the user is responsible for not fulfilling events that the target task has still not bound.''~\cite[\S4.4]{BSC:2019:OmpSs2}}
Otherwise, the task will run to completion but its dependencies will not be released before the event counter is decremented in the continuation callback \code{continue\_cb} (Lines~\ref{line:continue_cb1}--\ref{line:continue_cb2}).

In order to ensure that all continuations are eventually completed, a \emph{polling service} is registered with the OmpSs-2 runtime (Line~\ref{line:poll_mpi3}).
This polling service is regularly invoked by the runtime and---in the function \code{poll\_mpi} in Lines~\ref{line:poll_mpi1}--\ref{line:poll_mpi2}---calls \code{MPI\_Test} on the continuation request.\footnote{Note that OpenMP does not currently provide such an interface, which requires the user to create a \emph{progress task} or spawn a \emph{progress thread} that yield after testing.}
This will invoke all available continuations and eventually cause all tasks of the next timestep to become available for execution once the respective send and receive operations of the previous timestep have completed.

With only approximately 15 lines of code (including setup and tear-down), it is possible to integrate MPI communication with a task-parallel application to fully overlap communication and computation.
If adopted into the MPI standard, this approach is fully portable and allows an application to focus on application-level concerns while leaving MPI-level concerns such as request management to MPI.
By making the interaction between non-blocking MPI operations and the task scheduler explicit, this approach ensures that both MPI and the task programming system support all required operations (with the exception of strong progress guarantees, as discussed in \autoref{sec:rationale_progress}).
Moreover, we believe that integrating such an interface into MPI enables more efficient implementations as it allows direct access to the request structure instead of handling opaque request objects.

\section{MPI Continuations: Details}
\label{sec:rationale}
A scheme similar to the one proposed here could be implemented outside of the MPI library with an interface similar to TAMPI.
However, a main advantage of the integration with MPI is that the continuations can be invoked as soon as \emph{any} thread calls into MPI and determines the associated operations to be complete.
In more complex applications, this allows for the invocation of continuations set up by one part of the application during an MPI call issued in another part of the application, potentially reducing the time-to-release of blocked tasks and thus preventing thread starvation.

\subsection{Restrictions}
\label{sec:restrictions}
As stated earlier, continuations may be executed by application threads while executing communication functions in MPI.
Exceptions are \code{MPIX\_Continue[all]}, which may be called from within a critical region protected by a mutex it would then attempt to acquire again inside the continuation.
\autoref{lst:block_unblock_abt} provides an example where the call to \code{MPIX\_Continue} happens in a critical region guarded by a mutex, which is necessary to prevent signals from being lost due to the \code{unblock} callback being executed on another thread before the thread executing \code{block} started waiting in \code{pthread\_cond\_wait}.

While not prohibited, the use of blocking operations inside continuations should be avoided as it may cause a call to a nonblocking MPI procedure to block on an unrelated operation.
No other continuation may be invoked in MPI calls made from within a continuation to avoid stack overflows due to deep nesting of continuation calls.

By default, continuations may not be invoked from within signal handlers inside the MPI implementation as that would restrict the application to using only async-signal-safe operations.
MPI implementation relying on signals to interact with the network hardware should defer the execution to the next invocation of an MPI procedure by any of the application threads or hand over to a progress thread.
However, the application may signal that a callback is async-signal safe, as discussed in \autoref{sec:continue_info}.

\begin{listing}
\begin{lstlisting}[breaklines=true, postbreak=\mbox{\textcolor{red}{$\hookrightarrow$}\space}]
void block(thread_state_t *ts, MPI_Request *req) {
    int flag;
    pthread_mutex_lock(&ts->mtx);
    MPIX_Continue(&req, &flag, &unblock, ts, MPI_STATUS_IGNORE, cont_req);
    if (!flag) pthread_cond_wait(&ts->cond, &ts->mtx);
    pthread_mutex_unlock(&ts->mtx);
}
int unblock(MPI_Status *status, void *data) {
    thread_state_t *ts = (thread_state_t *)data;
    pthread_mutex_lock(&ts->mtx);
    pthread_cond_signal(&ts->cond);
    pthread_mutex_unlock(&ts->mtx);
}
\end{lstlisting}
\caption{%
Code to block and unblock a POSIX thread.
The \fncode{thread\_state\_t} structure contains a thread-specific mutex and conditional variable.
Locking the mutex in \fncode{unblock} is necessary to prevent signals from getting lost.%
}
\label{lst:block_unblock_abt}
\end{listing}

\subsection{State Transitions}
\begin{figure}
\centering
\includegraphics[width=.9\columnwidth]{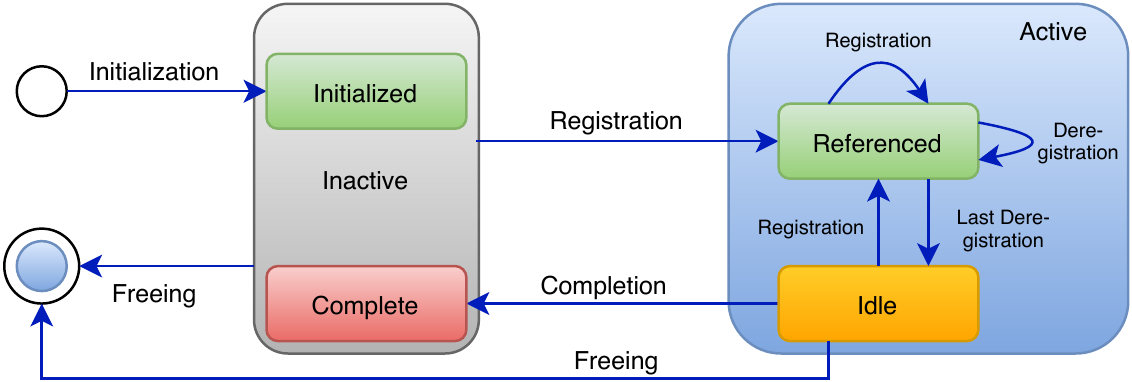}
\caption{State diagram of continuation requests.}
\label{fig:cont_state}
\end{figure}

Unlike existing request types in MPI, continuation requests (CR) do not represent a single operation but a set of operations.
Extending the semantics of persistent requests~\cite{Bangalore:2019:ECE}, \autoref{fig:cont_state} depicts the state diagram of CRs.
That state may change with every new registration or completion of a continuation.
A newly \emph{Initialized} or otherwise \emph{Inactive} CR becomes \emph{Active Referenced} when a continuation is registered.
It remains \emph{Active Referenced} if additional continuations are registered.
Upon completion of continuations, they are deregistered from the CR.
The CR becomes \emph{Active Idle} upon the deregistration of the last active continuation.
An \emph{Active Idle} CR may become either \emph{Active Referenced} again if a new continuation is registered or \emph{Complete} if a \emph{Completion} function such as \code{MPI\_Test} is called on it.
It is possible to call \code{MPI\_Request\_free} on an active CR, in which case the CR cannot be used to register additional continuations and will be released as soon as all previously registered continuations have completed.

A continuation may be attached to a CR itself and registered with a separate CR, allowing applications to build complex patterns of notifications.
The new continuation will then be executed once all continuations registered with the first CR have completed.

\subsection{Thread-safety Requirements}

Multiple threads may safely register continuations with the same continuation request in parallel but only one thread may test or wait for its completion at any given time.
This allows for continuations to be registered concurrently without requiring mutual exclusion at the application level.

\subsection{Progress Requirements}
\label{sec:rationale_progress}

The availability of continuations is only one piece of the puzzle to reliably and portably integrate MPI with task-based applications.
The issue of progress in MPI is long-standing and while some implementations provide internal mechanisms for progressing communication (e.g., progress threads) this is not behavior users can rely on in all circumstances.
Thus, applications are still required to call into MPI to progress outstanding communication operations and---with the help of the proposed interface---invoke available continuation callbacks.
This proposal does not change the status quo in that regard.

It is thus left to the application layer to make sure that communication is progressing.
This can be achieved by regularly testing the continuation request, which could be done either through a recurring task, a polling service registered with the task runtime system (as shown for OmpSs-2 in \autoref{lst:mpi_continue_ex}), or by relying on a progress thread inside MPI.
The interface presented here is flexible enough to allow for its use in a wide range of circumstances and it defers to the application layer to choose the right strategy based on the available capabilities.

\subsection{Controlling behavior through the  \code{info} object}
\label{sec:continue_info}
The \code{MPIX\_Continue\_init} function takes an \code{info} object that may be used to control certain aspects of the continuation request and the registered callbacks, respectively.
We propose the following info keys for controlling continuation request \emph{behavior}, which have proven useful during our experiments:

\begin{description}
\item[\code{mpi\_continue\_poll\_only}] Continuations registered with a CR created with this key set to \code{true} will only be executed once the CR is tested for completion, e.g., using \code{MPI\_Test}. This may be useful for ``heavy'' callbacks or callbacks that may block the executing thread in order to not disturb the execution of other parts of the application, e.g., by blocking threads used by another library. This info key provides control over the point at which callbacks are executed. 
The default is \code{false}.
\item[\code{mpi\_continue\_enqueue\_complete}] If this info key is set, then a call to \code{MPIX\_Continue[all]} will always return \code{flag = 0} even if the operation has completed immediately. The continuation will then be enqueued for later execution. 
This option may be useful for applications expecting large numbers of incoming messages with only limited time for handling them, instead choosing to defer their reception to a later point. Without this  key, the caller would have to have a mechanism for deferring the handling of completed operations itself. The default is \code{false}.
\item[\code{mpi\_continue\_max\_poll}] This key sets the maximum number of continuations that should be invoked once the continuation request is tested. This may be useful in cases where limited time should be spent on processing continuations at once, e.g., an application-level communication thread responsible for handling both incoming and outgoing messages. 
The application may then choose to limit the number of continuations to be handled at once in order to guarantee progress on outgoing messages as well. The default is \code{-1}, meaning unlimited. Setting both \code{mpi\_\-continue\_max\_poll = 0} and \code{mpi\_continue\_poll\_only = true} is erroneous as it would cause no continuation registered with this continuation request to ever be executed.
\end{description}

In addition to the info keys described above, we also propose two info keys controlling the \emph{context} in which the implementation may execute the registered continuations:

\begin{description}
\item[\code{mpi\_continue\_thread}] This key may be set to one of the following two values: \code{application} and \code{any}. The \code{application} value indicates that continuations may only be executed by threads controlled by the application, i.e., any application thread that calls into MPI. This is the default. The value \code{any} indicates that continuations may be executed by \emph{any} thread, including MPI-internal progress threads if available. Some applications may rely on thread-local data being initialized outside of the continuations callback or use callbacks that are not thread-safe, in which case the use of \code{any} would lead to correctness issues. This key has no effect on implementations that do not use a dedicated progress thread.
\item[\code{mpi\_continue\_async\_signal\_safe}] If this Boolean value is set to \code{true}, the application provides a hint to the implementation that the continuations are async-signal safe and thus may be invoked from within a signal handler. This limits the capabilities of the callback, excluding calls back into the MPI library and other unsafe operations. The default is \code{false}.
\end{description}

The list of info keys discussed above reflects the different dimensions in which applications may want to control the behavior of the continuations interface.
The \code{info} mechanism in MPI provides an ideal mechanism for controlling these different aspects, allowing users to adapt its behavior to their needs.

\subsection{Request Cancellation}

MPI provides the ability to cancel outstanding MPI operations using the \code{MPI\_Cancel} procedure.
While cancellation of send operations has been deprecated, it is still possible (and at times useful) to cancel requests representing receive operations, e.g., to clean-up pre-posted receive operations at the end of an application run.
Thus, applications need to be able to cope with canceled requests in continuation callbacks.
This can be achieved by allocating and passing status objects in calls to \code{MPI\_Continue[all]} and check for cancellation in the callback function, as demonstrated in \autoref{lst:req_cancel}.
This alleviates the need for explicit callbacks notifying about the cancellation of requests.

\begin{listing}
\begin{lstlisting}
void completion_cb(MPI_Status *status, void *data) {
  int is_cancelled;
  MPI_Test_cancelled(status, &is_cancelled);
  if (is_cancelled) {
    return;
  }
  /* perform regular actions of the callback otherwise */
  ...
}
\end{lstlisting}
\caption{Callback checking for cancellation of the request it is attached to.}
\label{lst:req_cancel}
\end{listing}

\section{Comparison with the MPI\_T Interface}
\label{sec:mpi_t}
The MPI\_T API will provide an interface for callback-based event notification in the upcoming version 4 of the MPI standard~\cite{Hermanns:2019:MPIT}.
This interface is meant for performance analysis tools to be notified about events on arbitrary MPI objects, e.g., communicators or window.
It would thus be possible to track the completion of operations using events on the respective communicators.
However, this would require the application (or some library) to keep a mapping between requests and the asynchronous activity related to them, which would be tedious and error-prone.

Since the interface is not limited to coarse-grain MPI objects such as windows and communicators, this interface could also be utilized to register completion events on individual requests.
However, this approach has several caveats that will be discussed below.

\subsection{Overhead}
In order to register a callback with individual operation requests using MPI\_T events, the user has to use three distinct function calls.
In a first step, an event handle is allocated for the specific request object using \code{MPI\_T\_EVENT\_HANDLE\_ALLOC}.
This binds a certain event (chosen from a set of events provided by the MPI implementation) to the request object and provides an event handle.
In a second step, a callback is registered with that event-registration object by calling \code{MPI\_T\_EVENT\_\-REGIST\-ER\_\-CALL\-BACK}.
Eventually, the event handle has to be released using \code{MPI\_\-T\_\-EVENT\_\-HANDLE\_FREE}.
For non-persistent requests, this has to be done during every completion callback to avoid memory leaks caused by dangling event handles.

Altogether, these three API calls are likely to incur a significant overhead for the registration of callbacks.
Unfortunately, the authors do not yet have access to a working implementation of the MPI\_T callback interface to test this hypothesis, which remains as future work.

\subsection{Reliability}
In its current form, the MPI\_T interface poses two major correctness challenges for it to be used for callback-based completion notification.
First, the current definition allows implementations to drop events if it finds that too many events have occurred.
While this may be acceptable for performance analysis (e.g., in the context of profiling tools) it is highly problematic for applications whose correctness relies on callbacks being delivered for each operation request.
Even though it is possible to be notified about the number of dropped events, the information on the particular completion will be lost.

Second, an inherent race condition exists in this interface since an operation may have completed between the time it was issued and the time the event callback has been registered.
It is unclear if the implementation would still be required to deliver the event to the application.
At the same time, the restrictions on immediate execution of continuations laid out in \autoref{sec:restrictions} would have to be obeyed. 

\subsection{Portability and Usability}
Currently, the MPI\_T interface does not provide any predefined events, making it impossible to rely on certain events for correctness.
While in the future the MPI standard may choose to mandate support for events for operation request completion, its detection would still be potentially tedious due to the weak binding using string comparisons.
In contrast, the interface proposed in this work provides a well-defined interface that applications can rely on.

However, mandating the existence of a specific event on requests with rather restrictive semantics seems to run counter to the spirit of the MPI\_T interface, which is to provide freedom to implementors to define events and their specific semantics.

Moreover, the MPI\_T interface does not provide the full capabilities of the continuations interface proposed in this work.
For example, in contrast to \code{MPIX\_Continueall} it is not possible to register callbacks for groups of requests.
Moreover, the MPI\_T interface does not provide a way to explicitly progress outstanding events and test or wait for outstanding registered callbacks, which our proposal accomplishes through continuation requests.

Last but not least, the MPI\_T interface includes a hierarchy of safety levels for callbacks, which comprises \code{MPI\_T\_CB\_\-REQUIRE\_NONE} (no restrictions on the set of MPI functions that may be called inside the callback), \code{MPI\_T\_CB\_REQUIRE\_MPI\_REST\-RICTED} (set of allowed MPI functions restricted to MPI\_T), \code{MPI\_\-T\_CB\_REQUIRE\_THREAD\_SAFE} (like the previous level and the callback must be thread-safe), and \code{MPI\_T\_\-CB\_REQUIRE\_ASYNC\_\-SIG\-NAL\_\-SAFE} (the callback must be async-signal safe).
This hierarchy, while useful for tools using the MPI\_T interface, does not provide sufficient control for applications.
Most importantly, this hierarchy does not allow for the definition of callbacks that are thread-safe and may invoke arbitrary MPI functions, which may be crucial for massively multi-threaded task-based applications employing this interface for task management purposes.
In contrast, the interface proposed in this paper allows for fine-grain control of these two (orthogonal) aspects through info key-value pairs, as described in \autoref{sec:continue_info}.

Overall, we believe that while the MPI\_T interface uses a \emph{similar} mechanism for the notification of arbitrary events, it is not suitable for the purpose of efficient operation completion notification as it only offers a subset of features the interface proposed here provides.

\section{Evaluation}
\label{sec:evaluation}
We will evaluate the use of continuations using the NPB BT-MZ benchmark (extended from earlier measurements presented in~\cite{Schuchart:2020:Fibers}), through an integration in \parsec{}, and an integration with a dynamic load-balancing scheme in ExaHyPE, an adaptive mesh refinement (AMR) package built on top of MPI and Intel Threading Building Blocks (TBB)~\cite{Samfass:2020:LTO}.

All measurements were conducted on a Hewlett  Packard Enterprise (HPE) Apollo system called \emph{Hawk}, which is installed at the Center for High Performance Computing Stuttgart (HLRS) in Stuttgart, Germany.\footnote{Detail can be found at \url{https://www.hlrs.de/systems/hpe-apollo-hawk/}. Last accessed February 11, 2021.}
The system is comprised of dual-socket nodes equipped with 128\,GB of RAM and AMD EPYC 7742 CPUs with a nominal frequency of 2.25\,GHz.
The nodes are connected through a enhanced 9D torus using the Mellanox InfiniBand HDR200 interconnect.
The configurations of the used software are listed in \autoref{tab:software}.

All data points represent the mean of at least five repetitions, with the standard deviation plotted as error bars in the graph.

\begin{table}
\renewcommand{\arraystretch}{1.1}
\caption{Software configuration.}
\label{tab:software}
\begin{tabular}{llp{.43\columnwidth}}
\toprule
\textbf{Software} & \textbf{Version} & \textbf{Configuration/Remarks} \\
\midrule
\ompi & \code{git-0dc2325} & \code{--with-ucx=...} \\
UCX & 1.9.0 & \code{--enable-mt} \\
GCC & 10.2.0 & \textit{site installation} \\
OmpSs-2 & 2020.11.1 & built as Clang plugin \\
Clang & \code{git-523fd9d03f} & includes fix for issue found in dependency handling \\
TAMPI & \code{git-c3bd734} & \emph{default} \\
\bottomrule
\end{tabular}
\end{table}

\subsection{Implementation}
We have implemented a proof-of-concept (PoC) of continuations within \ompi.\footnote{The PoC implementation is available at \url{https://github.com/devreal/ompi/tree/mpi-continue-master}. Last accessed February 11, 2021.}
We have provided micro-benchmarks as part of the initial paper~\cite{Schuchart:2020:Fibers} and refer interested readers to it for details.
Since then, we have added support for the info keys described in \autoref{sec:continue_info}, which should only have a marginal impact on the overhead of continuation registration and execution.
However, the integration into the MPI library has proven useful due to the direct access to the request object structures as we otherwise would have to build structures around the opaque request handles, potentially incurring additional overhead for their management.

\subsection{NPB BT-MZ}

\begin{table}[b]
\caption{Problem sizes in the NPB BT-MZ benchmark~\cite{NPB:2019:ProbSizes}.}
\label{tab:npb_problem_sizes}
\centering
\begin{tabular}{cm{.12\columnwidth}cm{.25\columnwidth}m{.14\columnwidth}}
\toprule
\textbf{Class} & \textbf{Zones} & \textbf{Iterations} & \raggedright \centering \textbf{Grid Size} ($x \times y \times z$) &  \centering\arraybackslash \textbf{Memory} \\
\midrule
D  & $32 \times 32$ & 250 & \centering$1632 \times 1216 \times 34$ & \raggedleft\arraybackslash 12.8\,GB \\
E  & $64 \times 64$ & 250 & $4224 \times 3456 \times 92$ & \raggedleft\arraybackslash 250\,GB \\
\bottomrule
\end{tabular}
\end{table}

In this section, we demonstrate the use of MPI Continuations in the context of the NAS Parallel Benchmark application BT-MZ, a multi-zone block tri-diagonal solver for the unsteady, compressible Navier Stokes equations on a three-dimensional mesh with two-dimensional domain decomposition~\cite{Wijngaart:2003:NPBMZ}.
Zones of different sizes are distributed across processes using a static load-balancing scheme, with the difference in zone sizes being up to $20\times$ between the smallest and largest zones~\cite{Haoqiang:2006:PCMZ}. 
In each timestep and for each local zone, five computational steps are involved: forming the right-hand side, solving block-diagonal systems in each dimension \code{x}, \code{y}, and \code{z}, and updating the solution.
We use two representative problem sizes---classes D and E---for which the input configurations are listed in \autoref{tab:npb_problem_sizes}.
The reference implementation in Fortran uses OpenMP work-sharing constructs to parallelize nested loops during the updates to all local zones before collecting and exchanging all boundary data with neighboring ranks at the end of each timestep.
OpenMP parallelization is done over the outermost loop, which in most cases is the smallest dimension \code{z}, with the notable exception of the solution in the \code{z} dimension itself.

We have ported the Fortran reference implementation to \CC and implemented two variations using task-based concurrency.\footnote{The different variants of NPB BT-MZ discussed here are available at \url{https://github.com/devreal/npb3.3}.}
The first variation uses tasks to overlap the computation of zones within a timestep in a fork-join approach in between boundary updates, replacing OpenMP parallel loops with task-generating loops, with their execution coordinated using dependencies.
This already allows for exploiting fine-grain concurrency within a single timestep.

The second variant extends this to also perform the boundary exchange inside tasks, including the necessary MPI communication, effectively minimizing the coupling between zones to the contact point dependencies.
This variant requires some support from the tasking library to properly handle communicating tasks.
We use TAMPI in combination with OmpSs-2 as well as detached tasks in OpenMP available from Clang in its current developer repository.
We have attempted to use \code{taskyield} within OpenMP tasks using the Clang implementation, which resulted in deadlocks due to the restricted semantics of task-yield in OpenMP~\cite{Schuchart:2018:TIT}.
The implementation using detached tasks spawns a progress thread that tests the continuation request before calling \code{usleep} to yield the processor.

Given the limited concurrency in the OpenMP work-sharing loops, we present results achieved using 2, 4, or 8 processes per nodes (PPN; 64, 32, and 16 threads per process, respectively).
All codes have been compiled with the flags \code{-O2 -march=znver2 -mcmodel=medium} using the Clang compiler.
Process binding has been achieved using \ompi's \code{--map-by node:PE=\${num\-threads} --bind-to core} arguments to \code{mpirun}.

\begin{figure*}
\begin{subfigure}{\columnwidth}
\centering
\includegraphics[width=\textwidth, page=6]{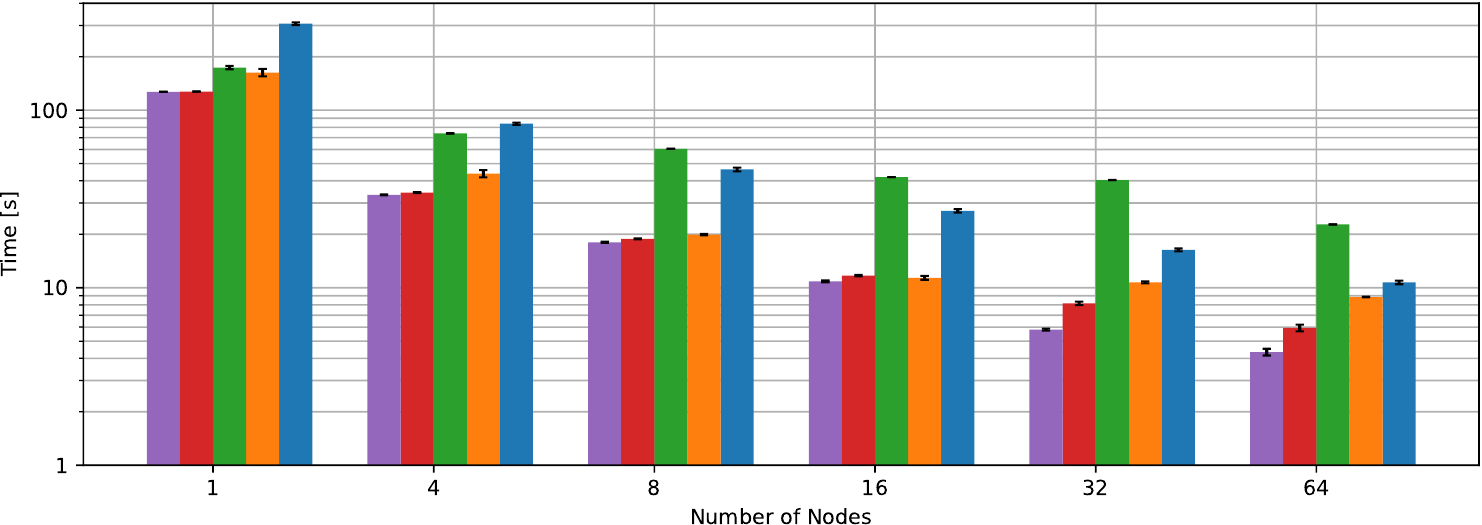}
\subcaption{$\text{PPN} = 2$}
\label{fig:npb_bt_mz_d:npn2}
\end{subfigure}
\hfill
\begin{subfigure}{\columnwidth}
\centering
\includegraphics[width=\textwidth, page=7]{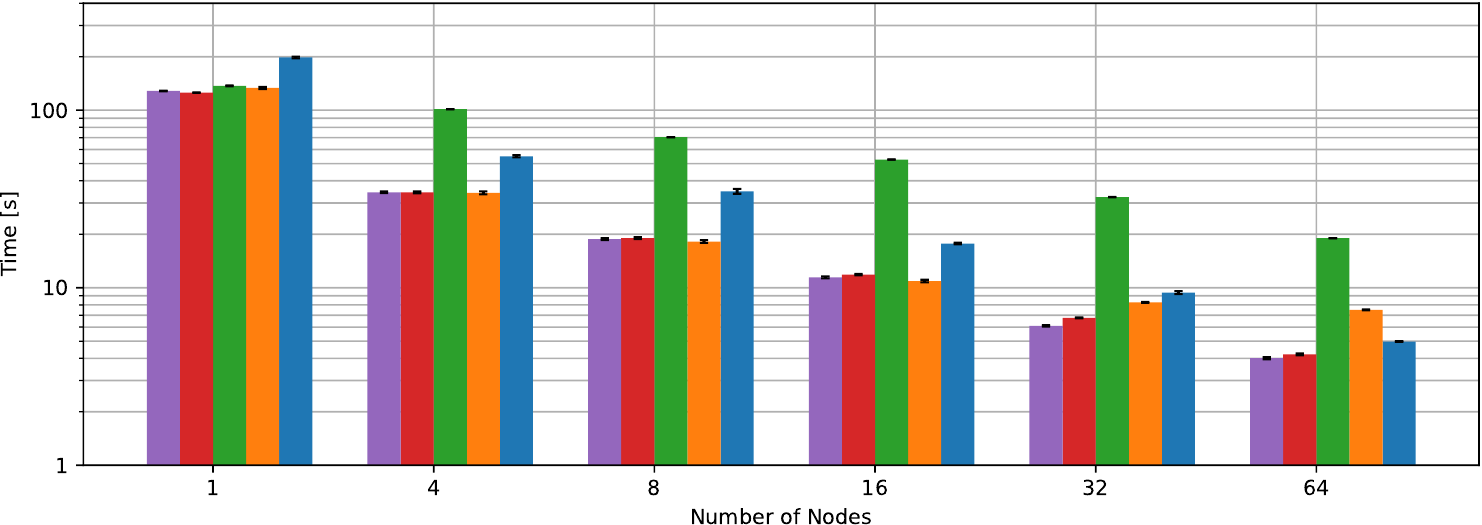}
\subcaption{$\text{PPN} = 4$}
\label{fig:npb_bt_mz_d:npn4}
\end{subfigure}

\begin{subfigure}{\columnwidth}
\centering
\includegraphics[width=\textwidth, page=8]{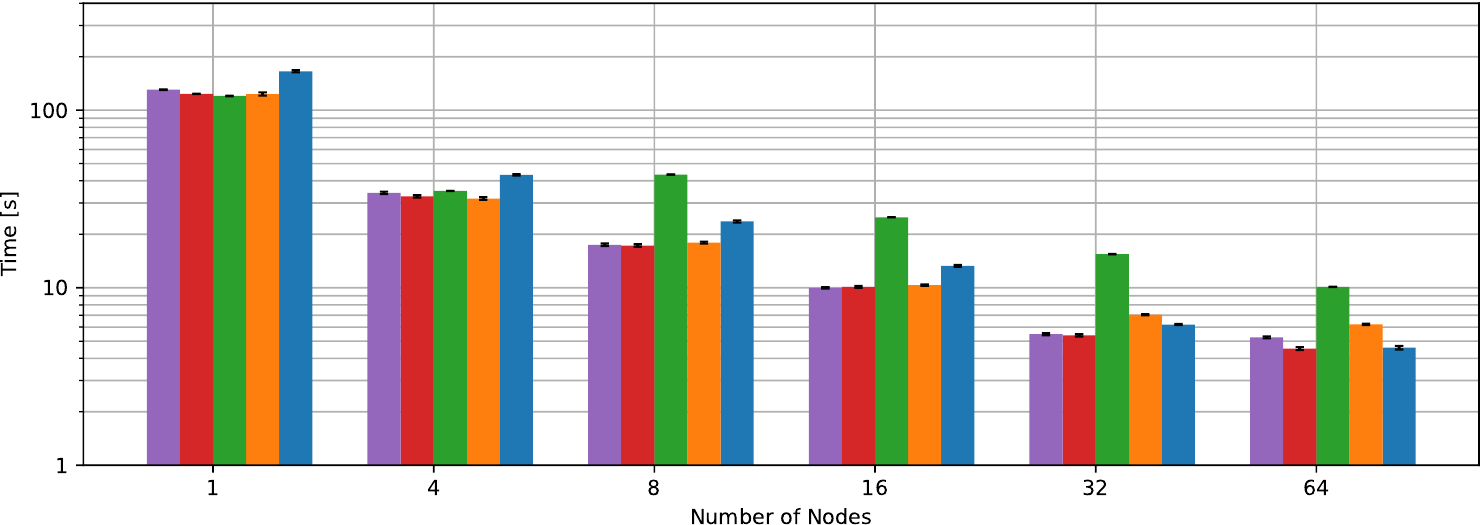}
\subcaption{$\text{PPN} = 8$}
\label{fig:npb_bt_mz_d:npn8}
\end{subfigure}
\hfill
\begin{subfigure}{\columnwidth}
\centering
\includegraphics[width=\textwidth, page=10]{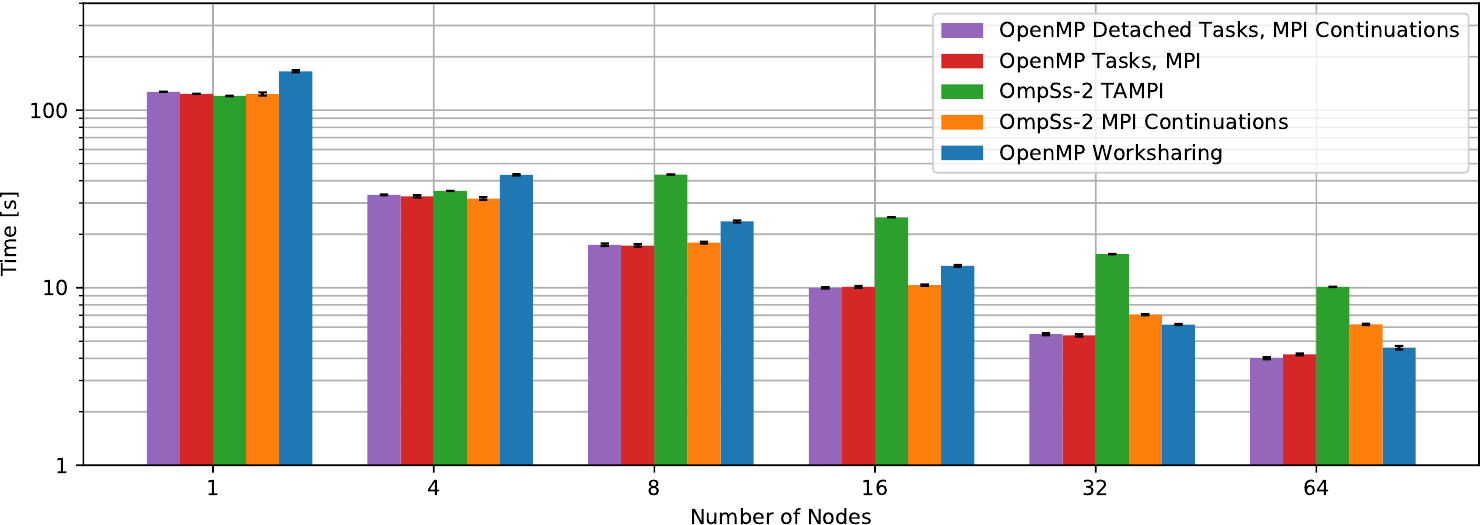}
\subcaption{Best runtime}
\label{fig:npb_bt_mz_d:best}
\end{subfigure}

\caption{NPB BT-MZ class D benchmark results.}
\label{fig:npb_bt_mz_d}
\end{figure*}

\paragraph{Class D}
\autoref{fig:npb_bt_mz_d} depicts the results for class D using 2, 4, and 8 processes per node, as well as the best runtime from each of the configurations.
Most notably, the OmpSs-2 variant using TAMPI yield significantly higher runtimes than the OpenMP variants, an effect that was reproduced multiple times for this problem size.
It is, however, notable that the OmpSs-2 variant using continuations yields lower runtimes than the variant using TAMPI, esp. for small values of PPN.
Similarly, the data shows that using continuations with OpenMP tasks yields better results for low PPN, esp. at the upper end of the process range (\autoref{fig:npb_bt_mz_d:npn2}).
For larger PPN the OpenMP variant using a task-based fork-join model achieves the same performance for higher PPN as the continuation-based variant with smaller PPN.
It should be noted, however, that at the upper end of the process range with $\text{PPN} = 8$, each process only receives two zones on average, diminishing the benefit of overlapping computations of multiple zones, esp. if the two zones share a boundary.

\paragraph{Class E}

\begin{figure*}
\begin{subfigure}{\columnwidth}
\centering
\includegraphics[width=\textwidth, page=6]{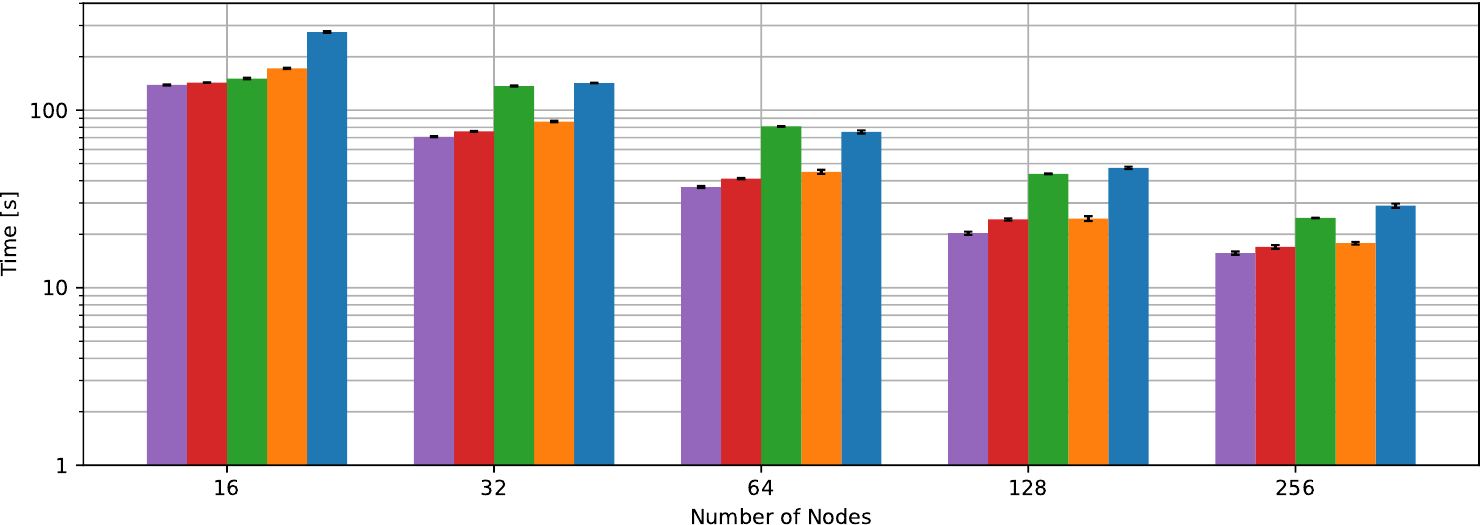}
\subcaption{$\text{PPN} = 2$}
\label{fig:npb_bt_mz_e:npn2}
\end{subfigure}
\hfill
\begin{subfigure}{\columnwidth}
\centering
\includegraphics[width=\textwidth, page=7]{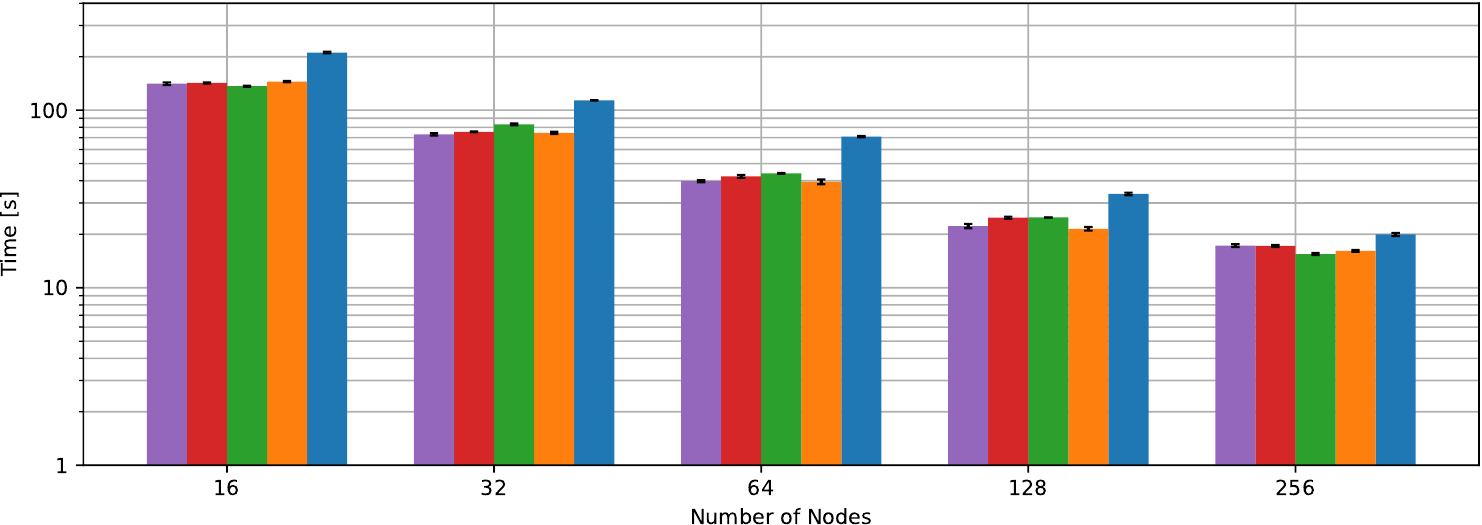}
\subcaption{$\text{PPN} = 4$}
\label{fig:npb_bt_mz_e:npn4}
\end{subfigure}

\begin{subfigure}{\columnwidth}
\centering
\includegraphics[width=\textwidth, page=8]{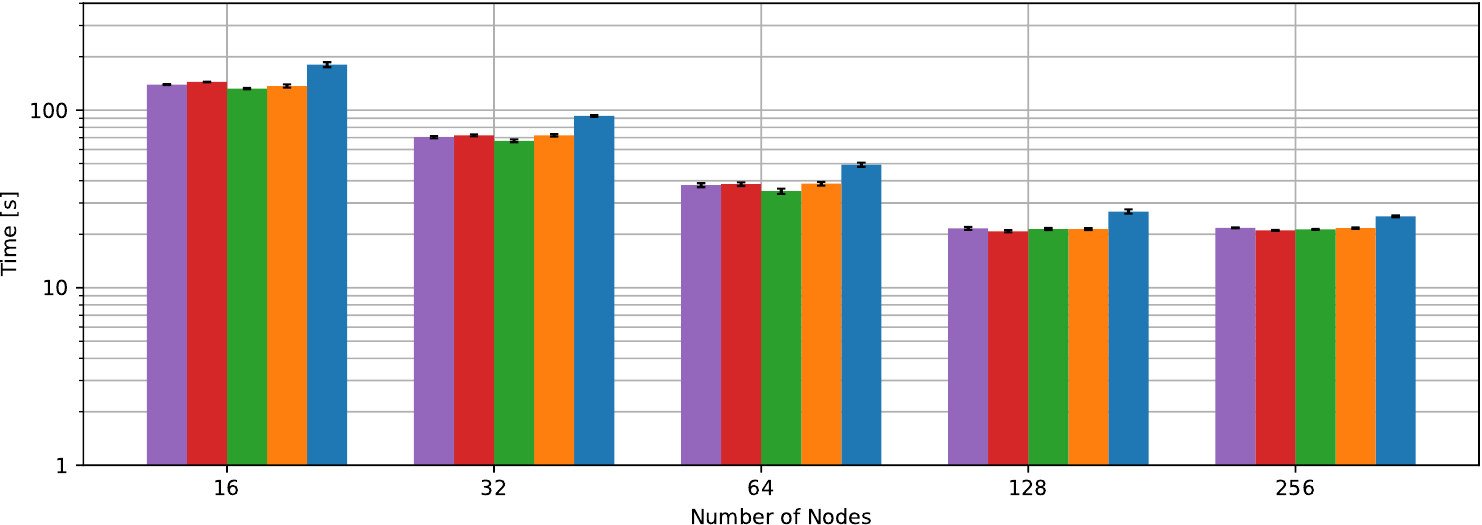}
\subcaption{$\text{PPN} = 8$}
\label{fig:npb_bt_mz_e:npn8}
\end{subfigure}
\hfill
\begin{subfigure}{\columnwidth}
\centering
\includegraphics[width=\textwidth, page=10]{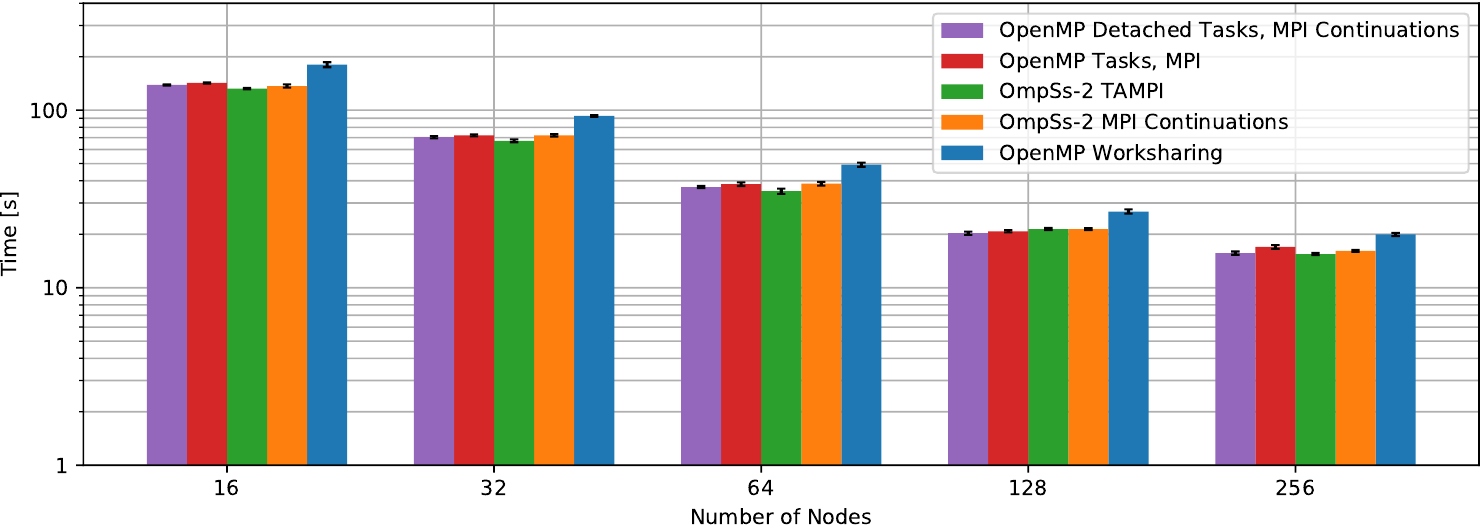}
\subcaption{Best runtime}
\label{fig:npb_bt_mz_e:best}
\end{subfigure}

\caption{NPB BT-MZ class E benchmark results.}
\label{fig:npb_bt_mz_e}
\end{figure*}

\autoref{fig:npb_bt_mz_e} shows the results for runs using class E, i.e., with $4\times$ more and larger zones than class D (\autoref{tab:npb_problem_sizes}).
Similar to class D, the scaling of the OmpSs-2 variant using TAMPI is inhibited for $\text{PPN} = 2$, in contrast to the other task-based variants.
This suggests that the handling of requests inside TAMPI and the required inter-thread synchronization harms performance especially for high numbers of threads.
However, this time for higher PPN the scaling is significantly better with TAMPI, at times slightly outperforming the OmpSs-2 variant using continuations (e.g., 32 and 64 nodes with $\text{PPN} = 8$).
The OpenMP variant using continuations again outperforms the fork-join variant for $\text{PPN} = 2$ and is mostly on par for other configurations.

Two factors may contribute to this observation.
First, with higher PPN, even with only 64 nodes there are merely eight zones per rank on average, with the static load balancing scheme potentially assigning only one or two zones to some of the ranks, diminishing the potential for overlapping computation and communication. 
Second, higher PPN also reduce the ability of MPI to offload communication to the network interface card and requiring more communication to be performed using node-local memory copies, requiring more work by the compute threads.

These two factors may explain the fact that the OpenMP variant using continuations performs best with small PPN values.
This in turn suggests, that the interface proposed in this paper can help improve the performance of task-based applications using MPI within tasks as lower PPN values may reduce the surface-to-volume ratio in some application and provide for low-overhead communication management even with higher numbers of threads.

\subsection{\parsec}

The \parsec{} tasking runtime~\cite{DAGuE:2011} provides an integrated task execution and communication environment, which tracks ownerships and moves data between nodes.
Originally proposed as a runtime for DPLASMA (a library of linear algebra algorithms and successor of ScaLAPACK)~\cite{Bosilca:2011:FDD,Choi:1992:ScaLAPACK}, PaRSEC has evolved into a generalized runtime with different frontends for applications that can be expressed as directed acyclic graphs.
At its core is a scheduler using POSIX threads to execute tasks and a global dependency tracking and communication system.
Communication itself is performed by a dedicated communication thread, allowing MPI to be used with \code{MPI\_THREAD\_FUNNELED} to avoid potential thread-synchronization overheads.

\begin{figure}
\centering
\includegraphics[width=.9\columnwidth, page=2]{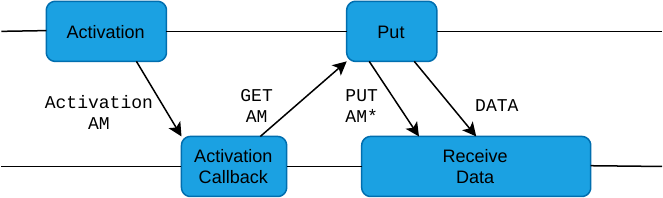}
\caption{The \parsec{} communication pattern for activating a dependency: an active message (AM) is sent to trigger the activation at the remote rank, which then sends an AM to request the data that flows along the dependency. With the two-sided MPI communication backend, the data communication itself involves an (internal) AM (marked with $^*$) to trigger the receive matching the send. The data communication happens without blocking the communication thread.}
\label{fig:parsec_comm_pattern}
\end{figure}

The communication thread thus handles both outgoing and incoming communication.
In its latest incarnation (on which this work is based), the interface has been redesigned to resemble a one-sided communication API providing \code{put}, \code{get}, and active-message capabilities, independent of the underlying communication backend.
In particular, the MPI backend emulates one-sided communication using two-sided send/recv by exchanging internal active messages (AM) that start a receive (\code{put}) or send (\code{get}) at the remote side before issuing the corresponding local operation.
On top of that, AMs are used to send activation messages for dependencies that have been resolved by completing a task.
The communication pattern involved in activating a dependency and transferring the data using an emulated \code{put} is depicted in \autoref{fig:parsec_comm_pattern}.

In the reference implementation, a small number of persistent receives are posted for each registered AM tag. 
In a tight loop, the communication thread sends outgoing messages and processes incoming messages.

Incoming activation messages require the communication thread to release dependent tasks and potentially trigger new communication operations to retrieve data required as input to the activated tasks.
This may require significant work on the part of the communication thread, in which time no other active message can be processed.

\begin{figure}
\centering
\includegraphics[width=.9\columnwidth, page=3]{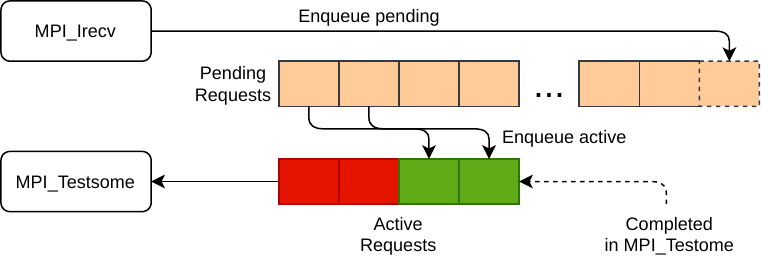}
\caption{Request management in the \parsec{} communication thread.}
\label{fig:parsec_comm_thread_request}
\end{figure}

As depicted in \autoref{fig:parsec_comm_thread_request}, the communication thread manages a fixed size array of requests that are passed to \code{MPI\_Testsome} to test for completion of communication operations, including send and receives posted for AMs and data messages.
This set of \emph{active requests} is kept small deliberately to mitigate the overhead of request checking in \code{MPI\_Testsome} and the subsequent reorganizing required once requests have completed.
This set is accompanied by a list of \emph{pending requests} that are posted send and receive operations and are eventually moved into the set of active requests once any active data message has completed.

While keeping the set of active requests small ensures efficient testing, it may lead to delays in completion detection: while a recently posted operation may have already completed, it may not be recognized as such until its request is moved into the set of active requests.

The communication thread also limits the number of concurrent outgoing data messages in order to prioritize the limited bandwidth on a first-come-first-serve basis (message prioritization based on urgency is done at higher levels inside PaRSEC).
This simple throttling mechanism prevents heavy contention and ensures that messages can be sent with a large fraction of the available bandwidth instead of hundreds of messages being transferred with a small share of the bandwidth, which would delay all messages.

\subsubsection{Implementation using Continuations}

The use of continuations in the context of the \parsec{} communication engine promised to provide faster reaction times to incoming active messages and completed transfers, by executing the action associated with an active message or completed communication.
We thus register continuations for all requests, essentially eliminating the set of \emph{pending requests} and the use of \code{MPI\_Testsome}.
Any time the MPI implementation finds operations to complete, the attached continuation would be enqueued for execution.
By default, continuations for \emph{any} of the operations would be eligible for immediate execution.

However, especially the activation AM callbacks may be expensive, as they may in turn trigger new communication operations.
Moreover, the communication thread has to potentially handle a large number of active messages in bursts, effectively prolonging the time until the it is able to handle outgoing messages again. 
Such bursts lead to poor performance as outgoing activations will be delayed, starving ranks depending on them.

We thus created different continuation requests for different classes of operations (incoming active messages as well as incoming and outgoing data messages) and---on the one used for AMs---set the info key \code{mpi\_continue\_\-poll\_only} discussed in \autoref{sec:continue_info} to \code{true} in order to restrict the execution of incoming active messages to the point when the communication thread calls \code{MPI\_Test} on the respective continuation request.
An exception are AMs used to emulate the one-sided communication described above, which are known to be of short duration since they only encompass posting a non-blocking send or receive.
We also set \code{mpi\_continue\_enqueue\_complete} to \code{true} for all AMs to enqueue continuations even if the receive is complete immediately, preventing the communication thread from having to process large number of incoming AMs in burst situations.

Similarly, we are restricting the execution of continuations of incoming data messages, as they may incur work in the scheduler (releasing tasks) or additional communication (forwarding messages in hierarchical collective communication operations).
However, the completion of an outgoing data message is typically a short operation and can thus be executed immediately, freeing up a slot for the next outgoing data message.

\subsubsection{QR Factorization}

\begin{figure}
\centering
\includegraphics[width=.9\columnwidth, page=9]{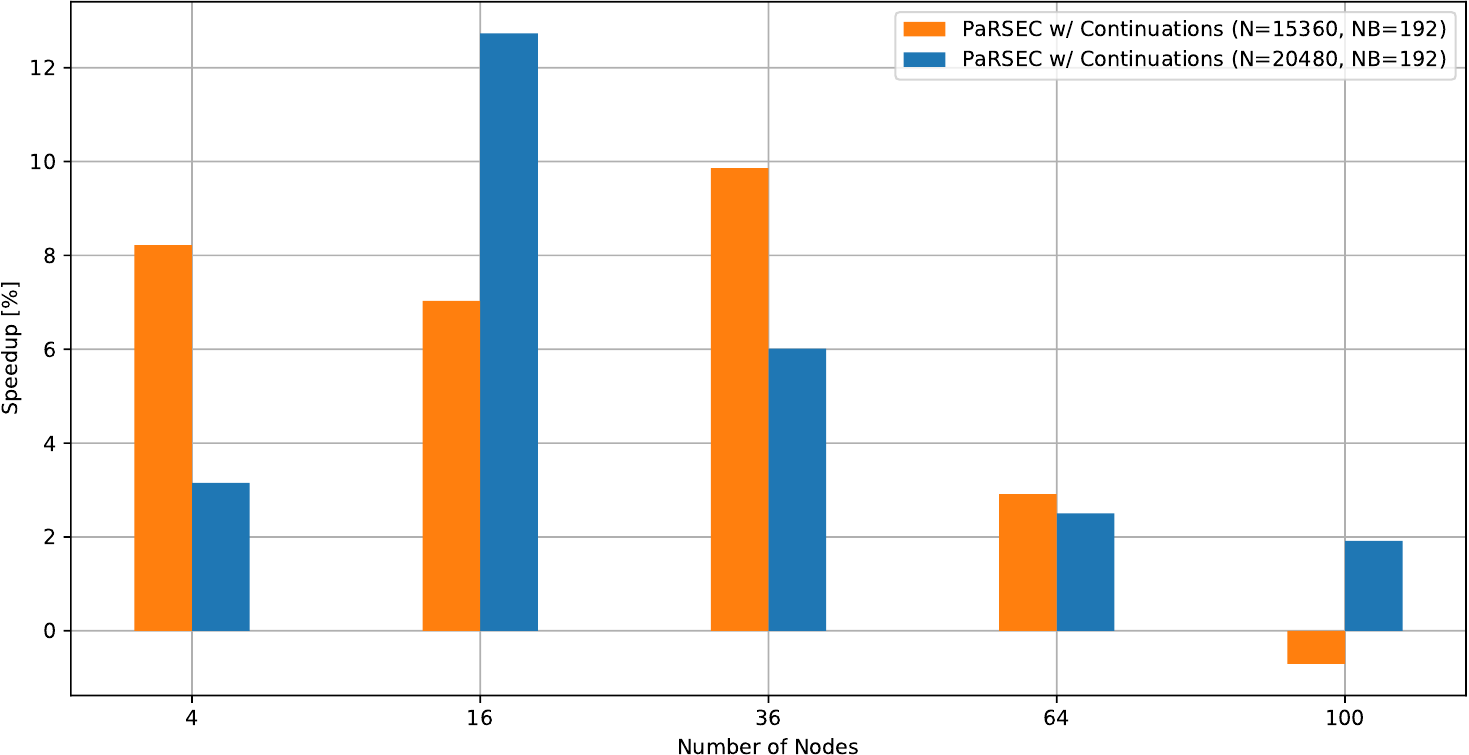}
\caption{Speedup of the QR factorization in DPLASMA using continuations over \parsec{}'s reference implementation with different per-node matrix and tile sizes.}
\label{fig:parsec_dgeqrf}
\end{figure}

The weak-scaling benchmark results for the QR factorization are depicted in \autoref{fig:parsec_dgeqrf}, using matrix sizes of $15\text{k}^2$ and $20\text{k}^2$ elements per node and a tile size of 192.
We found that the usage of continuations is most effective on smaller matrix sizes and relatively small tile sizes as that requires a fast exchange of both activations and data as worker thread starvation may occur otherwise.
Here, we observe more than 12\% speedup on 16 nodes ($\text{N}=20\text{k}^2$) and 10\% speedup on 36 nodes ($\text{N}=15\text{k}^2$), while for larger numbers of nodes the effect diminishes.
For larger tile sizes, we have observed neither a positive nor negative impact on the performance, suggesting that these configurations are less sensitive to latency and thus do not benefit from faster responses.

It is left as future work to further explore the use of continuations in other use-cases of \parsec{} and to analyze \parsec's communication requirement at higher node counts.

\subsection{ExaHyPE}

The ExaHyPE project has recently proposed a novel approach for load-balancing task-based adaptive mesh refinement (AMR) applications by offloading tasks from ranks with high loads to ranks with lower loads~\cite{Samfass:2020:LTO}.
Load balancing decision are made based on instrumenting MPI waiting times in at runtime.
They serve as input to a diffusive load balancing scheme in which purely local offloading decisions are made to mitigate the impact of overloaded processes, leading to improved load balance.
The project uses Intel Threading Building Blocks (TBB) to manage shared memory parallelism on top of MPI for distributed memory parallelism.
Due to its domain decomposition of the underlying space-tree created by tri-partitioning, ExaHyPE exhibits severe load imbalances at configuration other than 28 or 731 ranks.

\begin{figure}
\centering
\includegraphics[width=.8\columnwidth, page=1]{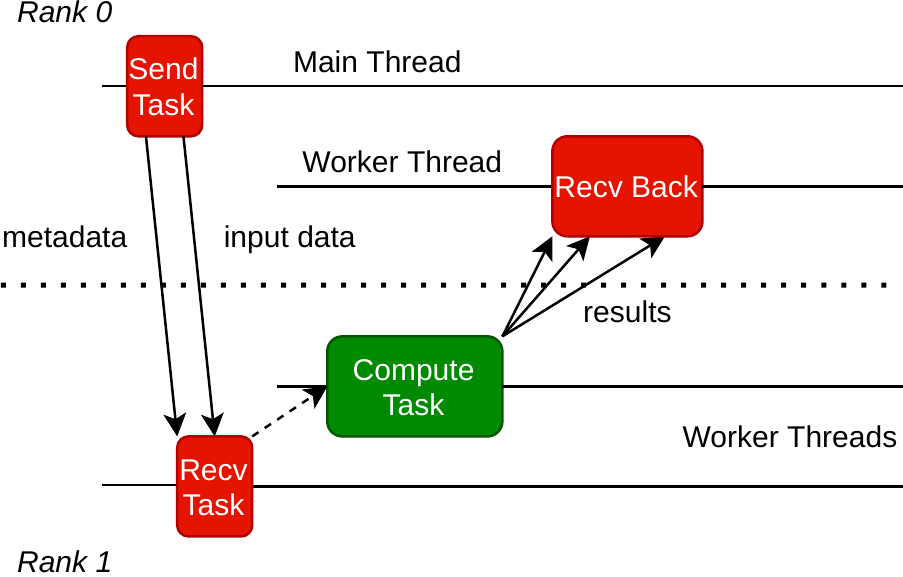}
\caption{Offloading a task in ExaHyPE.}
\label{fig:exahype_offloading}
\end{figure}

Tasks are being offloaded by processes identified as critical ranks to other processes based on the amount of time waited for a kick-off message to start the next iteration.
The decision on whether and where to offload a task is made at the time the task is discovered by the main thread during the traversal of the grid.
If local threads are idle,  tasks will be scheduled for local execution, i.e., they are not subject to being sent to another process. 
If, on the other hand, all local threads are busy and some additional tasks are available locally, the task will be sent to another rank, whose load was determined to be lower than the local load.
\autoref{fig:exahype_offloading} depicts the flow of an offloaded task.
Offloading a task consists in sending a short message containing the task metadata (e.g., timestep size and a timestamp) followed by a second message containing the task's input data, which are sent in two distinct messages as packing them into one would incur significant overhead. 

A small number of pre-posted persistent receives for metadata are used to identify incoming tasks.
Once such a receive completes, a receive for the task input data is initiated and upon completion the task is enqueued for execution by any of the worker threads.
As soon as a worker thread has completed the execution of an offloaded task, it initiates sending back the results, which consists of three messages.
At the offloading source, the non-blocking receives are posted in the callback signaling the completion of the send requests, which is useful for keeping the number of active requests low.

While it would be possible for the main thread to post the receives for the task results immediately after sending the task input data, we found that the extra time spent on posting the three receive operations induces significant overhead for the main thread (which is on the critical path) and thus slows down the overall execution.

In case the offloading target cannot return the result for the task in time, a so-called \emph{emergency} is triggered and the number of tasks offloaded to that rank is reduced.
If multiple emergencies occur, the rank will not receive any tasks for a number of timesteps and the algorithm tries to shift the load to other ranks.

All communication happens using non-blocking send and receive operations, which are managed by a central \emph{offloading manager} that invokes a callback upon detecting their completion, essentially implementing a form of continuations in the application space.
Since offloading a task involves multiple messages in both directions, it is necessary to manage groups of requests whose combined completion triggers the invocation of a single callback.
The offloading manager thus uses multiple parallel data structures to manage the mapping between MPI requests and callbacks.

Worker threads will poll the offloading manager for completed communication operations, passing a subset of the active requests to \code{MPI\_Testsome} (similar to \parsec{}).
However, it is not clear what constitutes an optimal number of requests to pass to the test function.
On the one hand, too little requests might cause some tasks that were offloaded to slow ranks and complete too late to block the detection of tasks returned from fast ranks, thus leading to unnecessarily large numbers of emergencies.
On the other hand, a large number of requests might incur significant overhead inside \code{MPI\_Testsome} during the linear walk through the request array to check all request for completion.
In order to ensure fast detection of completed tasks, the reference implementation opted for eagerly testing as many requests as possible while trying to avoid repeated reallocation of the request array.

Overall, this scheme is a natural fit for MPI Continuations.

\subsubsection{Implementation using Continuations}

In our implementation using MPI Continuations, we were able to replace much of the complexity of managing groups of requests with a single call to \code{MPIX\_Continueall}, reducing the amount of code necessary to implement the registration of callbacks with groups of requests, as shown in \autoref{tab:locs}.
An even more significant reduction in code can be found in the progression of registered requests.
Where the reference implementation has to assemble a continuous array of requests to pass to \code{MPI\_Testsome} and handle the returned array of indices, the use of continuations only requires invoking \code{MPI\_Test} on the continuation request.
In addition to the reduction in lines of code, we also removed the data structures that were dedicated to managing the required mapping between requests, groups, callbacks, and callback arguments in the application space (not shown in \autoref{tab:locs}).
Overall, the use of the continuations interface has reduced the overall complexity of a part of code that is central to the load balancing scheme described above.

Since requests do not have to managed in the application space, we have implemented a scheme that shifts the callback registration for task send requests to worker threads, allowing the use of a single continuation for send and receive requests.
This scheme has not yielded any improvements in the reference implementation, presumably due to the larger number of simultaneously active requests to be handled by the offloading manager.

\begin{table}[b]
\caption{Lines of code required for submitting and progressing requests in ExaHyPE with and without MPI Continuations.}
\label{tab:locs}
\centering
\begin{tabular}{l r r r }
\toprule
\textbf{Operation} & \textbf{Reference} & \textbf{Continuations} & $\Delta$ \\
\midrule
Submitting requests & 55 & 36 & $-19$ \\
Progressing requests & 133 & 17 & $-116$ \\
\bottomrule
\end{tabular}
\end{table}

\subsubsection{Results}

\begin{figure}
\centering
\begin{subfigure}{.75\columnwidth}
\includegraphics[width=\textwidth, page=9]{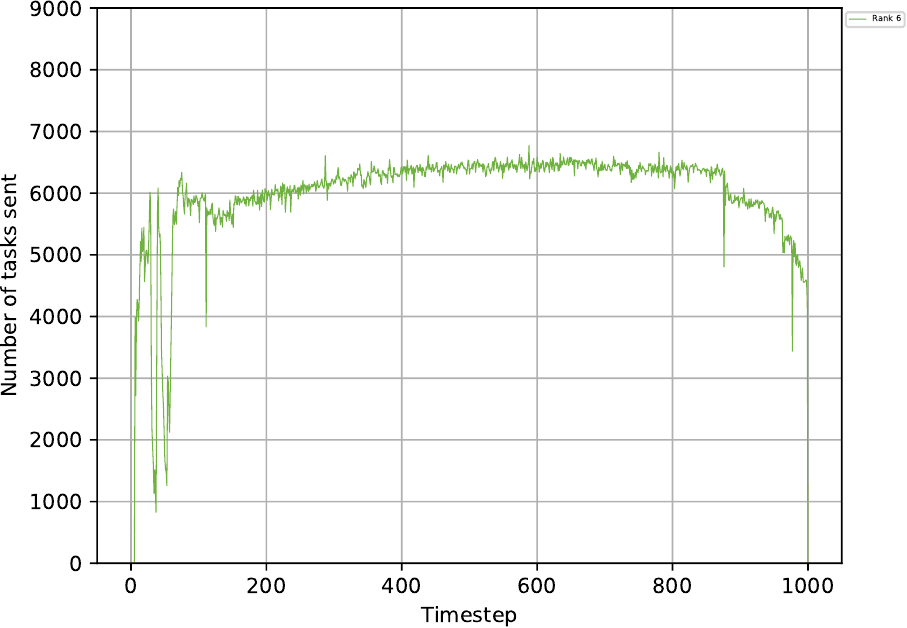}
\caption{Reference implementation.}
\label{fig:offloaded_tasks:reference}
\end{subfigure}

\begin{subfigure}{.75\columnwidth}
\includegraphics[width=\textwidth, page=9]{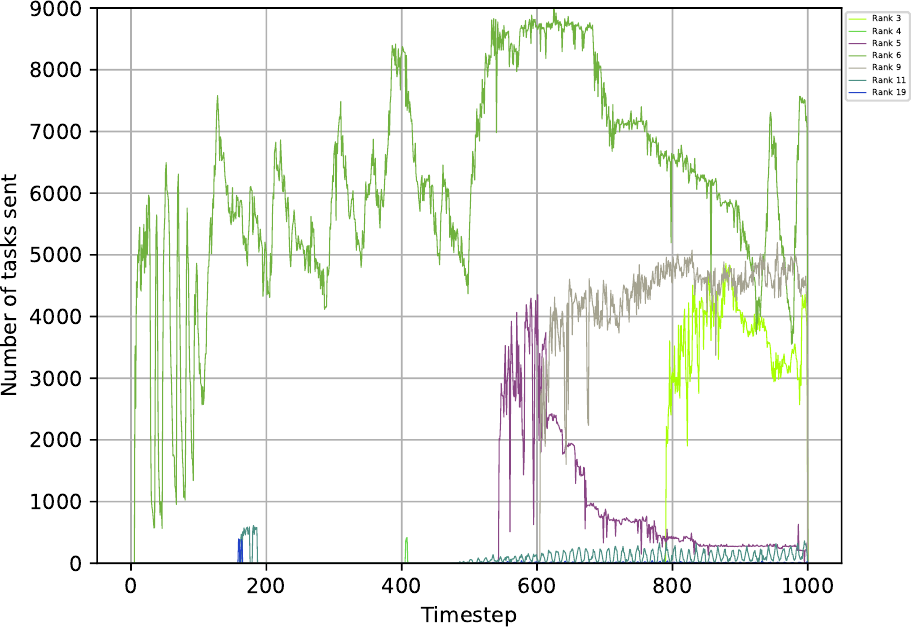}
\caption{Using continuations.}
\label{fig:offloaded_tasks:cont}
\end{subfigure}
\caption{Number of offloaded tasks over 1000 iterations. Each line represents a single offloading source. Using continuations results in higher numbers of tasks being offloaded and a shift in the critical rank.}
\label{fig:offloaded_tasks}
\end{figure}

In our benchmark, we use ExaHyPE's ADER-DG implementation to solve the compressible Navier Stokes equations for the simulation of clouds~\cite{Krenz:2020:Clouds}.
For small problem sizes, we have not observed significant differences between the reference implementation and the implementation using continuations.
%
However, \autoref{fig:offloaded_tasks} shows the number of tasks offloaded over a run of 1000 timesteps on a grid of $81^3$ points on 24 ranks with 32 threads each.
\autoref{fig:offloaded_tasks:reference} shows that in the reference implementation, only one rank is offloading tasks as that rank remains on the critical path, reaching a peak at approximately 6500 tasks.
In contrast, \autoref{fig:offloaded_tasks:cont} shows that when using continuations, the same rank is able to offload just short of 9000 tasks ($+35\%$), at which point other ranks are detected as being critical by the algorithm, causing them to offload tasks as well.

\begin{figure}
\centering
\begin{subfigure}{.75\columnwidth}
\includegraphics[width=\textwidth, page=1]{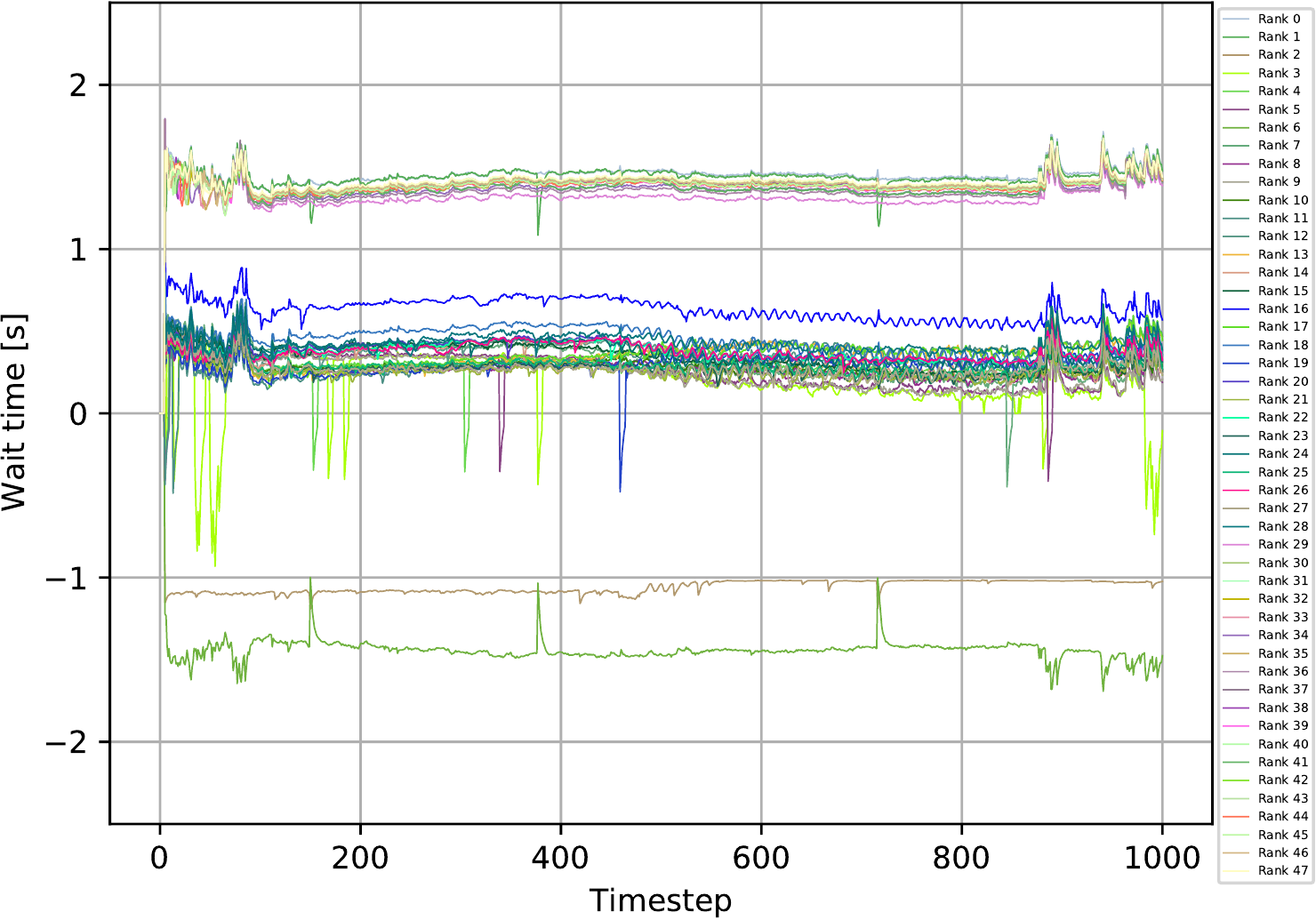}
\caption{Reference implementation.}
\label{fig:waittimes:reference}
\end{subfigure}

\begin{subfigure}{.75\columnwidth}
\includegraphics[width=\textwidth, page=1]{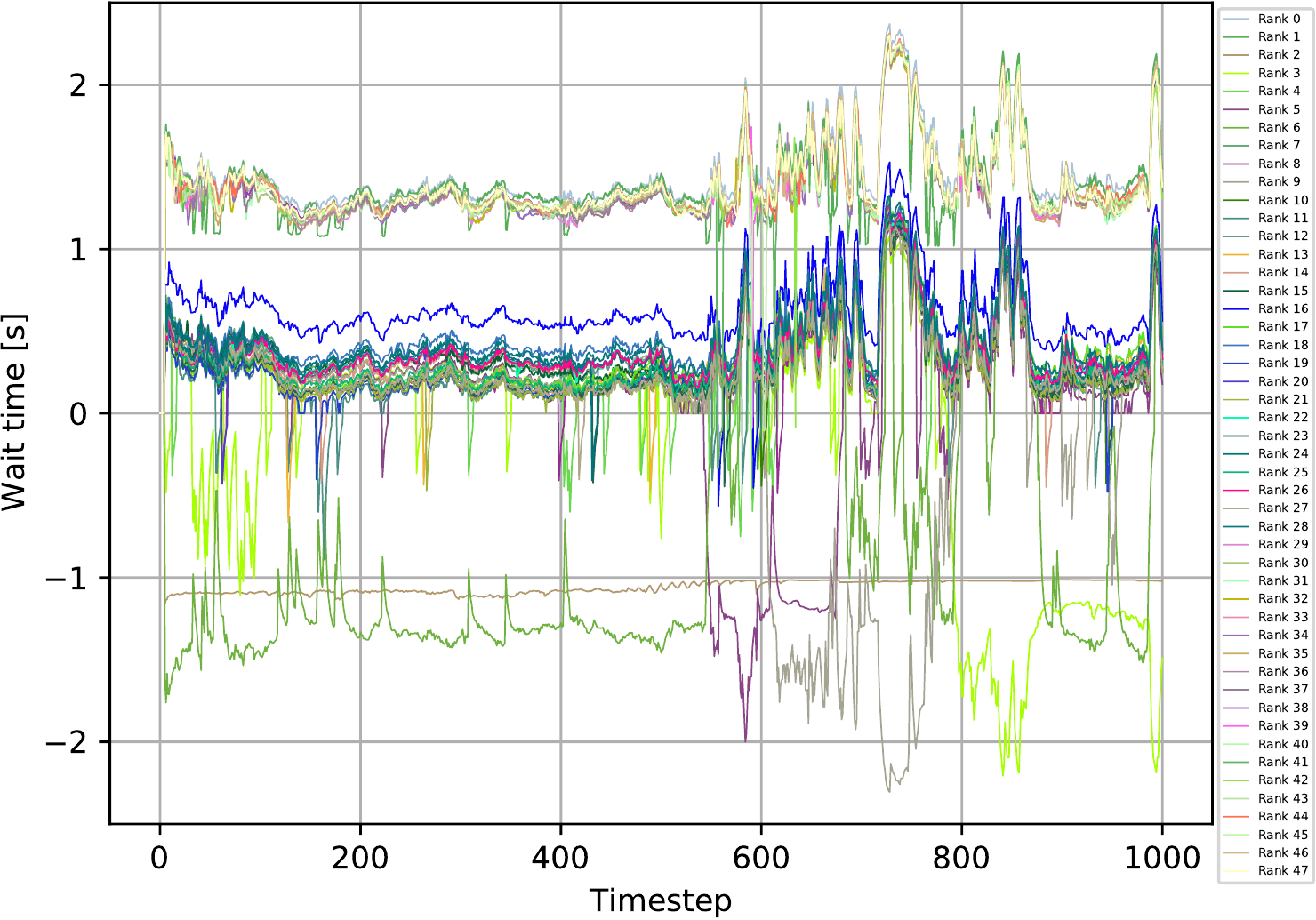}
\caption{Using continuations.}
\label{fig:waittimes:cont}
\end{subfigure}
\caption{Wait times per timestep over 1000 iterations on 48 ranks. Ranks with positive times wait for other ranks to complete the timestep. Ranks with negative wait times are being waited on. Using continuations yields lower runtimes on the critical path (lower half) until other ranks start offloading tasks, leading to disturbances.
}
\label{fig:waittimes}
\end{figure}

\autoref{fig:waittimes} shows the corresponding wait times for each iteration.
A positive wait time indicates that the rank is waiting for one or more other ranks (and thus could become an offloading target) while a negative wait time is used to indicate that the corresponding rank is being waited on (and thus should offload tasks).
In the reference implementation (\autoref{fig:waittimes:reference}), the rank on the critical path is being waited on for approximately 1.45\,s for the majority of timesteps.
With continuations, on the other hand, the wait time for the critical rank in the first 550 iterations is at 1.3\,s, which constitutes a 10\% speedup (\autoref{fig:waittimes:cont}).
However, as other ranks are detected as critical, a disturbance occurs because the offloading targets that have so far only processed tasks of the initial critical rank are now served tasks from other ranks as well, rendering them unable to process tasks in a timely manner.
The algorithm is able to identify this condition and adjust the number of tasks offloaded to recover from the overload.
Due to the diffusive nature of the algorithm, many more iterations will be required to achieve a stable state.
However, it is only through the use of continuations that enough tasks can be offloaded in this scenario to trigger the change in the critical rank. 


\section{Related Work}

Multiple efforts have been made in the past to improve specific aspects of the interaction between multi-threaded programming models and MPI~\cite{Hoefler:2010:EMS,Dinan:2013:EMI,Grant:2019:FPM}. 

Extended generalized requests have been proposed to encapsulate operations within requests and allow the MPI implementation to progress them through a callback~\cite{Latham:2007:EMG}.
Similar to the interface proposed here, extended generalized requests employs \emph{inversion of control} (IoC)~\cite{Mattsson:96:OOF}: application-level code is called by the MPI library in inversion of the usual call relation.
However, these requests are still polling-based and cannot be used for completion notification.

IoC is also used in the upcoming MPI\_T events interface, allowing applications and tools to be notified through callbacks about events from within the MPI library~\cite{Hermanns:2019:MPIT}.
However, the availability of events is implementation-specific and callbacks can only be registered for \emph{event types}, as opposed to specific operations.
The MPI\_T interface has been used to structure the interaction between MPI and task-based programming models, still requiring the mapping between MPI requests and task-information in the application layer~\cite{Castillo:2019:OCC}.

Continuations are not a new concept in the context of low-level communication libraries: UCX uses callback notification functions when initiating non-blocking operations~\cite{UCX:1.6}.
In contrast, other interfaces such as libfabric~\cite{OFI:2017}, Portals~\cite{Barrett:2018:Portals4}, and uGNI~\cite{Pritchard:2011:uGNI} rely on completion queues from which notifications are popped. 
While this would be a viable concept in MPI as well, we believe that continuations provide greater flexibility at the application level. 
Other low-level communication abstractions such as LCI~\cite{Dang:2018:LCI} provide lightweight completion notification by setting variables directly accessible by the application.

Another alternative would be a system similar to \emph{scheduler activations} in OS kernels~\cite{Anderson:1991:SAE}, where MPI would signal to some application-level entity that the current thread or fiber would block inside a call.
This has been proposed for the OpenShmem standard~\cite{Rahman:SCO:2020}.
However, its use is limited to tasking models that have strong rescheduling guarantees and would not work with OpenMP detached tasks.

\emph{Hardware-triggered operations} are being used to react to events in the network with low latency and to offload computation to the network interface card~\cite{Hoefler:2017:SHS,Schonbein:2019:INCA,Islam:2019:MUH}.
However, in-network computing is focused on operations on the data stream and likely cannot be utilized to execute arbitrary code in user-space, e.g., call into tasking libraries.
Exploring the boundaries of integrating these systems with the interface proposed here remains as future work.

The integration of tasking and fiber libraries into MPI has been proposed as a replacement for POSIX thread support to enable the use of blocking MPI calls inside tasks and fibers~\cite{Lu:2015:MUO}.
We have discussed this in detail in~\cite{Schuchart:2020:Fibers}.
Other proposals provide runtime-specific wrapper libraries around MPI, transforming blocking MPI calls to non-blocking calls and testing for completion before releasing a fiber or task~\cite{Stark:2014:EECS,Sala:2018:IIM,Sala:2019:TAMPI}.
These proposals fall short of providing a portable interface that can be used with arbitrary asynchronous programming models.

A proposal for completion callbacks for MPI request objects has been discussed in the context of the MPI Forum over a decade ago and rejected due to its broad focus~\cite{Hoefler:2008:RCCF}.
In contrast, the proposal outlined here avoids some of the pitfalls such as callback contention, ownership of requests, and unclear progress semantics.

A proposal made concurrently to ours has been made in the form of \code{MPI\_Detach}~\cite{Protze:2020:MPID}.
While similar in its goals and its general interface (registering completion callbacks for one or more requests), it differs in subtle but important ways.
Most importantly, the API does not provide immediate completion and uses a general progress procedure to process outstanding completion callbacks without the testing and waiting capability provided by continuation requests.
However, the existence of a concurrent proposal underscores the need for MPI to provide an interface for its integration with task-based programming models.

The interface proposed here is especially well-suited for coupling MPI with node-local asynchronous programming models such as Open\-MP~\cite{openmp5.0}, OmpSs-2~\cite{BSC:2019:OmpSs2}, and Intel TBB~\cite{Reinders:2007:TBB} or in combination with cooperatively scheduled fiber libraries such as Argobots~\cite{Seo:2018:AAL} or Qthreads~\cite{Wheeler:2008:QAA} to ease the MPI communication management burden users are facing.

\section{Conclusions}
\label{sec:conclusions}

In order to tackle the challenges of asynchronous activities communicating through MPI, we have proposed an extension to the MPI standard that allows users to register continuations with active operations that will be invoked once the MPI library determines the operations to be complete.
In this paper, we have extended this interface with fine-grained controls meant to steer the behavior of the MPI implementation to accomodate the needs of applications by using the info key mechanism of MPI.
Moreover, we have discussed early experiences in using this interface in combination with four different task-based runtime systems, namely OpenMP detached tasks, OmpSs-2, Intel TBB, and the \parsec{} custom scheduler.

The overhead of our proof-of-concept implementation is sufficiently low to not impact existing applications and its use shows similar or improved performance over existing approaches.
Overall, MPI continuations may provide a powerful tool for reducing the latency in thread-parallel message passing applications while reducing the programmer's burden on writing and maintaining complex software for managing MPI requests in the application space.

The interface has been designed with higher-level abstractions such as \CC futures for MPI in mind but it remains as future work to provide an actual implementation on top of MPI Continuations.
However, we are confident that our design will be useful in the design of an asynchronous \CC MPI interface.

\section*{Acknowledgments}

This research was in parts funded by the German Research Foundation (DFG) through the German Priority Programme 1648 Software for Exascale Computing (SPPEXA) in the SmartDASH project.
The research leading to these results has received funding from the European Union’s Horizon 2020 research and innovation programme under the ChEESE project, grant agreement No. 823844.
This material is based upon work supported by the National Science Foundation under Grant No. \#1664142 and the Exascale Computing Project (17-SC-20-SC), a collaborative effort of the US Department of Energy Office of Science and the National Nuclear Security Administration.

\bibliographystyle{elsarticle-num.bst}
\bibliography{references}

\end{document}